\documentclass[prl,twocolumn,superscriptaddress,showpacs]{revtex4}
\usepackage{graphicx}
\usepackage{amsmath}
\usepackage{amssymb}
\usepackage{dcolumn}
\usepackage{bm}
\usepackage{textcomp}
\usepackage{ulem}
\usepackage{colordvi}

\begin{document}

\title{Nuclear Charge Radii of $^{7,9,10}$Be and the one-neutron halo nucleus $^{11}$Be}

\author{W.~N\"ortersh\"auser}
\affiliation {GSI Helmholtzzentrum f\"ur Schwerionenforschung GmbH, D-64291 Darmstadt, Germany} %
\affiliation {Institut f\"ur Kernchemie, Universit\"at Mainz, D-55128 Mainz, Germany}

\author{D.~Tiedemann}
\affiliation {Institut f\"ur Kernchemie, Universit\"at Mainz, D-55128 Mainz, Germany}

\author{M.~\v{Z}\'{a}kov\'{a}}
\affiliation {Institut f\"ur Kernchemie, Universit\"at Mainz, D-55128 Mainz, Germany}

\author{Z.~Andjelkovic}
\affiliation {Institut f\"ur Kernchemie, Universit\"at Mainz, D-55128 Mainz, Germany}

\author{K. Blaum}
\affiliation {Max-Planck-Institut f\"ur Kernphysik, D-69117
Heidelberg, Germany}

\author{M.~L.~Bissell}
\affiliation {Instituut voor Kern- en Stralingsfysica, Katholieke Universiteit Leuven, B-3001 Leuven, Belgium}

\author{R.~Cazan}
\affiliation {Institut f\"ur Kernchemie, Universit\"at Mainz, D-55128 Mainz, Germany}

\author{G.W.F.~Drake}
\affiliation {Department of Physics, University of Windsor, Windsor, Ontario, Canada, N9B 3P4}

\author{Ch.~Geppert}
\affiliation {Helmholtzzentrum f\"ur Schwerionenforschung GmbH, D-64291 Darmstadt, Germany} %
\affiliation {Physikalisches Institut, Universit\"at
T\"ubingen, D-72076 T\"ubingen, Germany}

\author{M.~Kowalska}
\affiliation {CERN, Physics Department, CH-1211 Geneva 23, Switzerland}

\author{J. Kr\"amer}
\affiliation {Institut f\"ur Kernchemie, Universit\"at Mainz, D-55128 Mainz, Germany}

\author{A. Krieger}
\affiliation {Institut f\"ur Kernchemie, Universit\"at Mainz, D-55128 Mainz, Germany}

\author{R.~Neugart}
\affiliation {Institut f\"ur Kernchemie, Universit\"at
Mainz, D-55128 Mainz, Germany}

\author{R.~S\'anchez}
\affiliation {Helmholtzzentrum f\"ur Schwerionenforschung GmbH, D-64291 Darmstadt, Germany} %

\author{F.~Schmidt-Kaler}
\affiliation {Quanteninformationsverarbeitung, Universit\"at Ulm, D-89069 Ulm, Germany}

\author{Z.-C.~Yan}
\affiliation {Department of Physics, University of New Brunswick, Fredericton, New Brunswick,
Canada E3B 5A3}

\author{D.~T.~Yordanov}
\affiliation {Max-Planck-Institut f\"ur Kernphysik, D-69117
Heidelberg, Germany}

\author{C.~Zimmermann}
\affiliation {Physikalisches Institut, Universit\"at T\"ubingen, D-72076 T\"ubingen, Germany}

\date{\today}

\begin{abstract}
Nuclear charge  radii of $^{7,9,10,11}$Be have been
determined by high-precision laser spectroscopy. On-line
measurements were performed with collinear laser
spectroscopy in the $2s_{1/2} \rightarrow 2p_{1/2}$
transition on a beam of Be$^{+}$ ions. Collinear and anticollinear laser beams were used simultaneously
and the absolute frequency determination using a frequency comb
yielded an accuracy in the isotope-shift measurements of about
 1~MHz. Combination with accurate calculations of the
mass-dependent isotope shifts yield nuclear charge radii. The charge radius
decreases from $^7$Be to $^{10}$Be and then increases for
the halo nucleus $^{11}$Be. When comparing our results with
predictions of {\it ab initio} nuclear structure
calculations we find good agreement. Additionally, the
nuclear magnetic moment of $^7$Be was determined to be
$-1.3995(5)~\mu_{\rm N}$ and that of $^{11}$Be from a previous $\beta$-NMR measurement was
confirmed.
\end{abstract}
\maketitle

The discovery of halo nuclei in 1985 \cite{Tan85} triggered
a large number of experiments in order to understand the
properties and behavior of these extraordinary systems. In
a halo nucleus, individual nucleons - in most cases
neutrons - can reside far away from the nuclear core. The
best established and investigated neutron-halo nuclei are
$^{6,8}$He, $^{11}$Li, and $^{11}$Be. In a neutron-halo
nucleus, the change of the nuclear charge distribution
provides information on the interactions between the
different subsystems of the strongly clustered nucleus.
Such changes may originate from two effects: firstly, the
motion of the nuclear core relative to the center of mass,
and secondly core polarization induced by the interaction
between the halo neutrons and the core. In the case of
helium the nuclear core is tightly bound and only a 4\%
contribution of core excitation is expected \cite{Pie01}.
The dominant part of the change is induced by the
center-of-mass motion. Conversely, the $^{9}$Li core in
$^{11}$Li is rather soft and core polarization can play a
much more important role \cite{Neu08}. Separation between
these effects is not trivial since the correlation between
the halo neutrons needs to be considered. Thus, to develop
a consistent picture of neutron-rich light nuclei, more
data are necessary for different systems, which is a
motivation to determine accurately the charge radii of
$^{7,9,10,11}$Be. $^{11}$Be is the first one-neutron halo
nucleus for which the charge radius is reported.

The only known model-independent way to determine charge
radii of unstable isotopes is the laser spectroscopic
measurement of the nuclear-volume dependent isotope shift.
Nuclear-volume shifts for isotopes of the light elements
are very small, and short-lived isotopes can be produced
only in marginal quantities. Therefore, high-precision
measurements are extremely difficult. Moreover, it is a
major challenge to separate these small effects from the
mass shifts which dominate by three orders of magnitude. A
breakthrough on the theoretical side was the sufficiently
accurate calculation of mass shifts for systems with up to
three-electrons \cite{Yan00,Yan03,Puc06,Yan08,Puc08}. These
results have been used for the first time to determine the
charge radii of $^{6,7,8,9}$Li \cite{Ewa04} and $^{6}$He
\cite{Wan04}. Later, the charge radii of the two-neutron
halo nucleus $^{11}$Li \cite{San06} and recently that of the
four-neutron halo nucleus $^{8}$He \cite{Mue07} were
determined based on these calculations. The results are now
benchmark tests for nuclear structure theory.

{\it Ab initio} descriptions of the structure of light
nuclei based on realistic two- and three-nucleon
interactions between individual nucleons have been
developed during the past years. Such are Greens-Function
Monte-Carlo (GFMC), Fermionic Molecular Dynamics (FMD), and
Large Basis No-Core Shell Model (NCSM) calculations, which
have predicted nuclear charge radii of beryllium isotopes.
Moreover, the charge radius of $^{7}$Be gives constraints
for the determination of the $^7$Be$(p,\gamma)^8$B $S$-factor
that is crucial for solar neutrino experiments
\cite{Cso98,Nav06}. For these reasons, the isotope shift of
beryllium isotopes has recently attracted considerable
interest. Measurements on trapped and cooled ions in a Paul
trap have been proposed for the D1 line \cite{Zak06} and
first results in the D2 line were reported \cite{Nak06}.
However, these measurements have not yet reached the
accuracy required to extract the field shift.

We have developed a technique to determine the optical
transition frequencies in collinear laser spectroscopy and
used it for on-line measurements on $^{7,9,10,11}$Be. The
radioactive isotopes were produced at ISOLDE (CERN) with a
1.4~GeV proton beam impinging on a uranium carbide target.
Beryllium atoms were ionized with the laser ion source
(RILIS). After extraction, acceleration to 50~kV and mass
separation, the ions were delivered to the collinear laser
spectroscopy setup \cite{Neu81}. This has been used for
isotope-shift measurements on many elements down to neon
($Z = 10$) \cite{Gei00}. However, for very light elements
the systematic uncertainties caused by the measurement of
the acceleration voltage exceed the nuclear volume effect.
Now, we can overcome these limitations employing a
frequency comb and measuring the absolute transition
frequencies $\nu_p$ for parallel and $\nu_a$ for
antiparallel geometry of ion and laser beams. This yields
the rest frame frequency $\nu_0$ independently of the
acceleration voltage via the relativistic formula
$\nu_0^2=\nu_p \cdot \nu_a$.

\begin{figure}[tbp]
\includegraphics[width=\columnwidth]{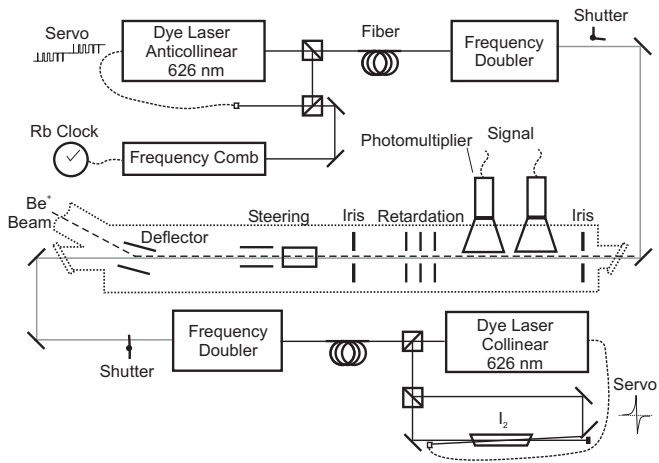}
\caption{Setup for collinear laser spectroscopy with parallel
and antiparallel excitation and a frequency comb for reference.} \label{fig:setup}
\end{figure}

A schematic layout of the setup and the laser system is
shown in Fig.~\ref{fig:setup}. The ion beam is
superimposed with the co- and counter-propagating laser beams
by an electrostatic 10$^\circ$ deflector. The laser beams enter and leave the beamline through a pair of Brewster windows. Two adjustable irises and further sets of
fixed size apertures down to 5~mm diameter are used along
the beam line to ensure good overlap and to avoid stray
light in the optical detection region. Doppler tuning is
performed by changing the potential of the detection region
in the range of $\pm10$~kV. Two photomultipliers are used
for resonance fluorescence detection.


The output of two dye lasers was frequency-doubled to
produce ultraviolet light at 313~nm and the UV beams were
well collimated over a distance of about 8~m with a beam
diameter of about 3-4~mm, well adapted to the ion beam
size. To avoid strong optical pumping and saturation
broadening of the induced transitions, the UV light was
attenuated to less than 5~mW. The fundamental light at
626~nm for collinear and anti-collinear excitation was
frequency-stabilized in different ways. One of the dye
lasers was locked to an iodine line using frequency
modulation saturation spectroscopy, while the second one
was locked to a frequency comb (Menlo Systems FC1500).
During the beamtime, the frequency of the iodine-locked
laser was repeatedly measured with the frequency comb. In
total, 12 different iodine lines were used and their
frequencies measured with standard deviations on the order
of 20~kHz. The Rb clock that was used as a reference
oscillator for the frequency comb introduces an additional
systematic uncertainty of about 350~kHz. However, this
contribution cancels out for the isotope shifts in which
frequency differences are evaluated.

\begin{figure}[tbp]
\includegraphics[width=\columnwidth]{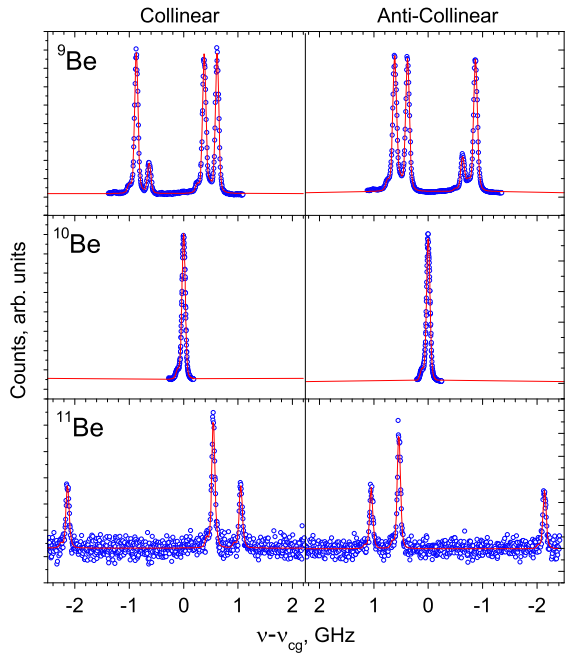}
\caption{(Color online) Fluorescence spectra
(\Blue{$\circ$}) for $^{9,10,11}$Be$^+$ in the $2s_{1/2}
\rightarrow 2p_{1/2}$ transition as a function of the
Doppler-tuned frequency in collinear (left) and
anti-collinear (right) excitation. Frequencies are given
relative to the respective hyperfine-structure center of
gravity for the odd isotopes and the resonance frequency
for $^{10}$Be. Lines (\Red{$-$}) are fitting results for
Voigt profiles as discussed in the text.}
\label{fig:spectra}
\end{figure}

Measurements were performed with the frequency of the
collinear laser locked to a suitable iodine line. The
voltage of the detection region was tuned to record the
hyperfine-structure pattern of the collinear excitation.
Then, the frequency of the anti-collinear dye laser was
locked to the frequency comb and its frequency chosen such
that the resonance pattern was covered by the same voltage
range. Repetitive scanning was performed by tuning the
voltage across this range. Dwell times of 20~ms per step
were used, resulting in about 4~s per scan. Two
remote-controlled mirrors were used to block alternately
one of the two laser beams during each scan. Typically $2
\times 50$ scans were accumulated for isotopes with high
yields, whereas up to $2 \times 400$ scans were taken for a
spectrum of $^{11}$Be (see Fig.~\ref{fig:spectra}).

The spectra were fitted with Voigt profiles of common
widths for all hyperfine structure components. The
hyperfine pattern was reproduced by direct calculation of
the $F\rightarrow F^{\prime }$ resonance position from
\begin{equation}
\nu_{FF^{\prime}}=\nu_{\rm cg}+ \frac{1}{2} \left[ A_{2p}~C(F,I,J)-A_{2s}~C(F^{\prime},I,J)\right]
\end{equation}%
with the center-of-gravity frequency $\nu_{\rm cg}$, the
$A$ factors of the ground ($A_{2s}$) and excited states
($A_{2p}$), respectively, and $C(F,I,J)=F\left( F+1\right)
-I\left(I+1\right) -J\left(J+1\right)$.
Chi-square-minimization was done by varying $A$ and $\nu
_{\mathrm{cg}}$ amongst other parameters. Weak satellite
peaks arising at higher acceleration voltages due to
inelastic collisions with residual gases were also included
in the fitting function. The achievable accuracy was tested
in an off-line beamtime and systematic shifts of the
transition frequency evaluated. Only a small effect caused
by beam misalignments could be observed.


The obtained isotope shifts $\delta\nu_{\rm IS}^{9,A} =
\nu_0(^A\mathrm{Be})- \nu _0(^9 \mathrm{Be})$ are listed in
Tab.~\ref{tab:ChargeRadii}. The quoted uncertainty
represents the standard error of the mean of individual
results from all measurements on a particular isotope. An
uncertainty of 0.5~MHz accounting for possible
misalignments of the laser and ion beam overlap was added
quadratically. Recoil corrections in the isotope shift are
only at the 10\% level of the final uncertainty. Charge
radii are deduced by subtracting the mass-dependent isotope
shift of the respective isotope pair as calculated in
Ref.~\cite{Yan08}. The remaining nuclear-volume shift
provides the change in the mean-square nuclear charge radius $\delta
\left\langle r_{\rm c}^2\right\rangle$ between two
isotopes. Absolute charge radii $r_{\rm c}$ must be related
to at least one isotope for which the absolute radius is
known and can then be deduced according to
\begin{equation}
r_{\rm c}^2(^A{\rm Be})=r_{\rm c}^2(^9{\rm
Be})+\frac{\delta\nu_{\rm IS}^{9,A}-\delta\nu_{\rm
MS}^{9,A}}{C} \label{eq:ChargeRadius}
\end{equation}
with the theoretically calculated electronic factor
$C=-16.912$~MHz/fm$^2$ \cite{Yan08}.

\begin{table}
\caption{\label{tab:ChargeRadii}%
Isotope shifts $\delta\nu_{\rm IS}^{9,A}$ in the D1 line and theoretical mass shifts
$\delta\nu_{\rm MS}^{9,A}$ \cite{Yan08,Puc08} for $^A$Be - $^9$Be. Uncertainties for the absolute charge radius $r_{\rm c}$ include the uncertainty in the reference radius $r_{\rm c}(^9{\rm Be}) = 2.519(12)~{\rm fm}$ \cite{Jan72}.}
\begin{ruledtabular}
\begin{tabular}{rccccccc}
 & $\delta\nu_{\rm IS}^{9,A}$, MHz & $\delta\nu_{\rm
MS}^{9,A}$, MHz &
$\delta \left\langle r_{\rm c}^2\right\rangle^{9,A}$, fm$^2$ & $r_{\rm c} $, fm \\
\hline
$^7$Be    & -49~236.9(9)  & -49~225.75(4) &   0.66(5) & 2.647(17) \\
$^9$Be    & 0             & 0               &           & 2.519(12) \\
$^{10}$Be & 17~323.8(13)  &  17~310.44(1)   &  -0.79(8) & 2.357(18) \\
$^{11}$Be & 31~565.0(9)   &  31~560.31(6)   &  -0.28(5) & 2.463(16) \\
\end{tabular}

\end{ruledtabular}
\end{table}

For stable $^{9}$Be, $r_{\rm c}$ was determined by elastic
electron scattering \cite{Jan72}, and by muonic atom
spectroscopy \cite{Scha80}. Reported charge radii of
2.519(12)~fm \cite{Jan72} and 2.39(17)~fm differ by 0.13~fm
but agree within the rather large uncertainty of the muonic
atom result. We have used the electron scattering result,
but we note that this was not obtained in a completely
nuclear-model independent way. Hence, reanalysis of the Be
scattering data, as performed for the proton \cite{Sic03},
would be useful to determine whether the small uncertainty
of 0.012~fm is reliable. However, a small change in the
reference radius causes primarily only a parallel shift of
all charge radii.

Results for $\delta \left\langle r_{\rm c}^2\right\rangle$
and $r_c$ are listed in Table~\ref{tab:ChargeRadii}. Isotope
shifts in the D2 line have also been measured and are still
under evaluation. The extracted charge radii agree with the
values reported here, but are less accurate due to
unresolved hyperfine structures.

\begin{figure}[tbp]
\includegraphics[width=\columnwidth]{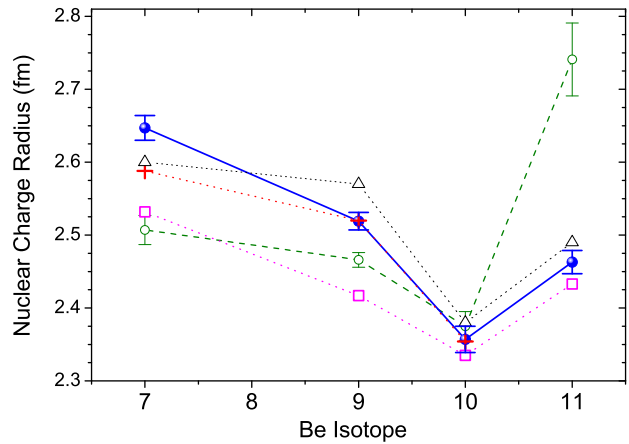}
\caption{(Color online) Experimental charge radii of
beryllium isotopes from isotope-shift measurements
(\Blue{$\bullet$}) compared with values from interaction
cross-section measurements (\Green{$\circ$}) and
theoretical predictions: Greens-Function Monte-Carlo
calculations (\Red{+}) \cite{Pie01, Pie02}, Fermionic
Molecular Dynamics ($\triangle$) \cite{Tor08}, {\it ab
initio} No-Core Shell Model (\Magenta{$\Box$})
\cite{For05,Nav06,Nav08}. } \label{fig:radii}
\end{figure}

The derived nuclear charge radii are shown in
Fig.\,\ref{fig:radii}: $r_{\rm c}$ decreases from $^7$Be to
$^{10}$Be, but then increases for $^{11}$Be. The decrease
is probably caused by the clusterization of $^7$Be into an
$\alpha$- and a triton-cluster, whereas $^{9,10}$Be are
considered to be $\alpha+\alpha+n$ and $\alpha+\alpha+n+n$
systems, respectively, and are more compact. According to a
simple frozen core two-body model the increase from
$^{10}$Be to $^{11}$Be can be attributed to the motion of
the $^{10}$Be core relative to the center of mass. Using $
r^2_c (^{11}{\rm Be})=R_{\mathrm{cm}}^2+r^2_{\rm c}
(^{10}{\rm Be})$ a rms distance of 7.0~fm between the
neutron and the center of mass of $^{11}$Be can be
extracted directly from $\delta \left\langle r_{\rm
c}^2\right\rangle$.

To test nuclear-structure theories, predictions from
different models are included in Fig.~\ref{fig:radii}.
Reported point-proton radii $r_{pp}$ were converted to
nuclear charge radii $r_{\rm c}$ by folding in proton
$R_{\rm p}$ \cite{Sic03} and neutron $R_{\rm n}$
\cite{Kop97} rms charge radii and adding the Darwin-Foldy
term \cite{Fri97} :
\begin{equation}
\left\langle r_{\rm c}^2\right\rangle = \left\langle r_{\rm
pp}^2\right\rangle + \left\langle R_{\rm p}^2\right\rangle
+ \frac{N}{Z}\left\langle R_{\rm n}^2\right\rangle +
\frac{3\hbar^2}{4m_{\rm p}^2c^2}~.
\end{equation}

Nuclear-model dependent charge radii extracted from
interaction cross-section measurements \cite{Tan88} show a
similar trend as our data but changes are smaller for the
isotopes below $^{10}$Be and overestimated for $^{11}$Be.
All theoretical model predictions show good agreement with
our results. Depending on the model, specific features are
especially well reproduced. In the case of GFMC
calculations \cite{Pie01,Pie02} the agreement for
$^{9,10}$Be is striking whereas the decrease between $^7$Be
and $^9$Be is slightly underestimated. Unfortunately, a
charge radius calculation for $^{11}$Be has not been
obtained to date. FMD \cite{Tor08} and improved NCSM
\cite{For05,Nav06,Nav08} calculations include the halo
nucleus and the change between $^{10}$Be and $^{11}$Be is
very well predicted. Concerning the other isotopes FMD does
a better job for the isotope pair $^{9,10}$Be, whereas the
change in the pair $^{7,9}$Be is best described by the NCSM
results (note that the absolute charge radius depends on
the reference radius).

\begin{table}
\caption{\label{tab:magmom} Magnetic dipole coupling constants $A$ of the $2s_{1/2}$ and the $2p_{1/2}$ states and the magnetic moments of the odd isotopes.}
\begin{ruledtabular}
\begin{tabular}{clllc}
Isotope   & $A_{2s}$ (MHz)     & $A_{2p}$ (MHz) & $\mu (\mu_{\rm N})$ & Ref. \\
\hline
$^7$Be    & -742.90(25)        & -140.17(18)    & -1.3995(5)   & this         \\
          &                    &                & -1.398(15)   & \cite{Kap98} \\
$^9$Be    & -624.97(4)       & -118.00(4)     &              & this                    \\
          & -625.008837048(10) & -118.6(36)    & -1.177432(3) & \cite{Win83,Ita83}\\
$^{11}$Be & -2677.4(8)       & -505.4(5)    & -1.6813(5)   & this           \\
          &                    &                & -1.6816(8)   & \cite{Gei99}   \\
\end{tabular}
\end{ruledtabular}
\end{table}

From fitting the observed hyperfine structures we obtain
the magnetic dipole constant $A$ for the $2s$ and the
$2p_{1/2}$ states as listed in Tab.~\ref{tab:magmom}. Based
on the precision measurement of $A_{2s}$ in $^{9}$Be
\cite{Win83} and the corresponding nuclear magnetic moment
$\mu=-1.177432(3)\,\mu_{\mathrm{N}}$ \cite{Ita83}, we
determine the magnetic moments of $^7$Be and $^{11}$Be. For
$^{11}$Be we confirm the previously measured value from an
optical pumping $\beta$-NMR experiment \cite{Gei99}. For
$^7$Be, an earlier value from optical hyperfine
measurements \cite{Kap98} is improved by more than an order
of magnitude.

To summarize, we have measured the charge radii of
$^{7,9,10,11}$Be with on-line frequency-comb based
collinear laser spectroscopy. From a simple frozen-core
two-body model we obtain a rms distance of about 7~fm
between the halo neutron and the center of mass in
$^{11}$Be. Comparison with elaborate nuclear structure
calculations (GFMC, FMD, and LBSM) shows in all cases good
agreement with our measurement. The magnetic moment of
$^{11}$Be has been confirmed and the accuracy for $^7$Be
considerably improved. The developed experimental technique
has the potential to be applied to the very short-lived
nuclei $^{12,14}$Be and can generally improve the accuracy
of collinear laser spectroscopy where this is required.

\begin{acknowledgments}
This work was supported by the Helmholtz Association
(VH-NG-148), BMBF (06TU263I, 06MZ215, 06UL264I), and the EU
(FP-6 EU RII3-CT-2004-506065). M.K. was supported by the EU
(MEIF-CT-2006-042114). We gratefully acknowledge technical
support by M.~Fischer and R.~Holzwarth from Menlo Systems
and by the ISOLDE technical group.

\end{acknowledgments}

\end{document}